\documentclass[pra,10pt,superscriptaddress,footnoteinbib]{revtex4}
\usepackage{amsmath}
\usepackage{latexsym}
\usepackage{amssymb}
\usepackage{graphics,epstopdf}
\usepackage[colorlinks=true, citecolor=blue, urlcolor=blue]{hyperref}\usepackage{epsf,graphics,graphicx}
\usepackage{placeins}

\newcommand{\n}{\noindent}

\newcommand{\ed}{\end{document}}
\newcommand{\beq}{\begin{equation}}
	\newcommand{\eeq}{\end{equation}}

\begin{document}\title{Hybridization and Field Driven Phase Transitions in Hexagonally Warped Topological Insulators  }
	\author{Anirudha Menon}
	\affiliation{Department of Physics, University of California, Davis, California 95616, USA}\email{amenon@ucdavis.edu}
	\author{Debashree Chowdhury}
	\affiliation{Department of Physics, Harish Chandra Research Institute, Chhatnag Road, Jhusi, Allahabad, U. P. 211019,India}\email{debashreechowdhury@hri.res.in}
	\author{Banasri Basu}
	\affiliation{Physics and Applied Mathematics Unit, Indian Statistical Institute, Kolkata 700108, India}
	\email{sribbasu@gmail.com}

\graphicspath{{C:/Newproj/}}
%%\vspace*{1cm}
\begin{abstract}
\n
In this paper we discuss the role of material parameters and external field effects on a thin film topological insulator(TI) in the context of quantum phase transition(QPT). First, we consider an in-plane tilted magnetic field and determine the band structure of the surface states as a function of the tilt angle.  We show that the presence of either a hybridization term or hexagonal warping or a combination of both leads to a semi-metal to insulator phase transition which is facilitated by their ${\cal PT}$ symmetry breaking character. We then note that while the introduction of an electric field does not allow for this QPT since it doesn't break ${\cal PT}$ symmetry, it can be used in conjunction with a tunneling element to reach a phase transition efficiently. The corresponding critical point is then non-trivially depend on the electric field, which is pointed out here. Then, we demonstrate that including a hexagonal warping term leads to an immediate ${\cal PT}$ symmetry violating QPT.
\end{abstract}
\maketitle

\section{introduction}
Topological insulators\cite{Kane05} are a new type of material that exhibit various distinctive surface properties. These materials are insulators in the bulk but possess topologically protected conducting surfaces. Unlike commonly known crystalline solids and magnets, which are associated with broken symmetries, topological insulators are unique in the sense that their edge properties are highly robust against disorder(non-magnetized) or fluctuations\cite{Hasan09,Kane10}. It is noted that the nature of the surface states depends on the dimension of the TI\cite{tahir, tahir10,tahir11,tahir12,tahir13} and these topologically protected states in two-dimensional and three-dimensional band insulators possess a large spin-orbit coupling. In addition, for each momentum on the Fermi surface of the edge state, the spin has a fixed defined direction transverse to the momentum which makes the edge states helical, revealing the Rashba spin-momentum coupling as another interesting property of the TIs. These unique features of TI's are of great significance in fundamental condensed matter physics and consequentially, TIs have become promising candidates for applications in spintronics and quantum computation\cite{Hasan09}. While there are proposals for the generation of pure spin current or novel magnetoresistance effects on purely theoretical grounds, defect induced bulk conductivities inhibit their experimental validation. The large bulk conductivity is common to most of the TI materials due to unavoidable imperfections in the composition and crystal structure, but the contribution of the bulk conductivity can be reduced relative to that of the surface states' contribution if the TI samples are prepared in the form of ultrathin films.

It has subsequently been shown that ultrathin TI films have some characteristic features \cite{11zuin,12zuin} which are not exhibited in the bulk samples, making the physics of TI in the ultrathin limit rich and interesting. More significantly, when the thickness of the film becomes comparable to the penetration depth of the helical surface states, the top and bottom states of a TI begin to hybridize leading to a crossover from 3D to localized 2D (surface) states, which can be easily accounted for by adding a small hybridization term to the TI Hamiltonian. On the other hand, a hexagonal warping term has been proposed recently to explain the experimentally observed 2D energy contours of the surface states of the topological insulator like $Bi_{2}Te_{3}$. In this class of TI’s the corresponding Fermi surface exhibits a snowflake like structure for large values of chemical potential \cite{Fu}. Fu confirms that to maintain the bulk topological invariance, there should exist out-of-the-plane components of the spins to conserve the net value of Berry’s phase, and this out of plane component of the Berry phase can only be explained in the presence of a hexagonally deformed cone of the TIs. Incorporation of this hexagonal warping term provides new possibilities to connect with the spintronics application based studies in TIs\cite{menon, jalil}.  It turns out that the hybridization term describing the surface state interaction also plays a very crucial role in explaining the quantum phase transition(QPT) in presence of a parallel magnetic field \cite{Burkov}, wherein the coupling between hybridization term and magnetic field term gives rise to the semimetal to insulator phase transition. There are also reports \cite{Burkov, menon} on the study of ultrathin topological insulator film with hybridization between the top and bottom surfaces, subjected to perpendicular magnetic field and it has been shown \cite{Burkov} that for an undoped film, a QPT occurs due to the competition between the Zeeman and the hybridization energies. This phase transition leads to a state with a quantized Hall conductivity equal to $e^2/h$ from a zero conductivity state. Moreover, when the Fermi level of any thin film TI is in the hybridization gap, there appears a  non-zero diamagnetic response due to the presence of a parallel magnetic field leading to a QPT from an insulator with a diamagnetic response to a semi metal \cite{zyuin1103}. In this context, magneto-optical response and quantum spin Hall and anomalous  Hall effects, and possible exitonic super-fluidity have also been studied theoretically.

Another important aspect of thin film TI's is their topological response to external fields and sensitivity to the strength of hybridization. The results of using in plane \cite{Burkov} and perpendicular magnetic field \cite{menon} on the thin film TIs are quite known. But how the tilted magnetic field affects the surface states as well as the behaviour of SdH oscillation pattern changes with the tilt angle, is quite interesting and is worth studying.  It is demonstrated in \cite{Liu}, that the external electric field can lead to QPT in TI's, which motivated us to introduce a perpendicular electric field in addition to an in-plane magnetic field in our present work. We've also considered the effects of hexagonal warping and a varying tunneling strength between the two surfaces along with the introduction of the EM fields, leading to interesting displays of phase transitions driven by symmetry breaking and changes topological invariants from bulk to surface.  

The organization of the paper is following:
In Sec. II, the ultrathin TI under consideration is in the presence of a tilted in-plane magnetic field, and we plot the dispersion to indicate semi-metal QPT's regime for different values of warping and hybridization. Sec. III contains two subsections, where the effect of electric field on the surface states of a warped thin TI is considered, in presence of zero and non-zero hybridization. In Sec. III A, we discuss the modified Hamiltonian in the presence of a perpendicular electric field and in the absence of hybridization and state the effects of including a QPT inducing warping term. In Sec. III B, we examine the effect of an external electric field driven QPT and demonstrate the critical relationship between the electric and magnetic fields and the hybridization piece at the point of phase transition. These effects are also considered under varying values of each parameter and on turning on a warping contribution. Conclusions are presented in Sec. IV.

\section{Hamiltonian and Energy Dispersion in presence of tilted magnetic field}

In this section, we start out by considering a thin warped TI film in a tilted in-plane magnetic field. The field is assumed to make an angle $\phi$ with the $x$ axis and lies in the $x-y$ plane. In this case, the Hamiltonian of an ultrathin TI (short sided in the $z$ direction) can be expressed as
\beq H_0 ( \vec{k} ) = v_{F} \tau^{z} ( \hat{z} \times \vec{\sigma}) .  (\vec{k} - k_{B_x} \tau^{z} \hat{y} -  k_{B_y} \tau^{z} \hat{x}) + \Delta_t \tau^x , \eeq

with $k_{B_i}= (\frac{eB_i d}{2c} - \frac{eB_i}{2mv_Fc})$, with $i=x,y$, containing the Aharaonov-Bohm and Zeeman contributions in each direction respectively. Here $d$ is the width of the sample in the z direction. In the limit that $d> 1$~nm~\cite{Burkov}, the Zeeman term maybe neglected and this will be assumed henceforth. This Hamiltonian respects ${\cal PT}$ symmetry in the absence of the hybridization term and the magnetic field. The introduction of the intersurface tunneling element is equivalent to adding "mass" to a Dirac Fermion, and hence it opens up a band gap. Note that the stabliity of edge state electrons to ${\cal PT}$ violating contributions and field driven topological disorder will be the central theme of this paper.
\newline
\newline
We now need to account for the magnetic field contribution to this Hamiltonian, and through the Pierels substitution  we have $\vec{k} \rightarrow \vec{k} -(\frac{e}{c \hbar}) \vec{A}$. Here $A$ is the gauge potential corresponding to the external magnetic field. In the present case, we can have  $\vec{B} = (B Cos \phi , B Sin \phi, 0)$ and using the Landau gauge to get $\vec{A} = (Bz Sin \phi , - B z Cos \phi, 0)$. So then,

\beq k_x \rightarrow k_x - \frac{eBd}{2c} Sin \phi \tau^z = k_x - k_{B_x} \tau^z, \quad  k_y \rightarrow k_y + \frac{eBd }{2c}Cos \phi \tau^z =  k_y + k_{B_y} \tau^z ,\eeq

where $\hbar$ has been set equal to 1. The $\tau^z$ terms in the above expressions represent the idea that we consider states localized near the top and bottom surfaces and hence the only possible values are $z$ are $\pm d/2$, implying that $z= (d/2) \tau^z$. The hexagonal warping term given by the Hamiltonian component

\beq H_w(\vec{k}) = \frac{\lambda}{2} (k_+^3 + k_-^3) \sigma^z ,\eeq

where $ k_{\pm} = k_x \pm ik_y$.
Adding the Hamiltonians, we arrive at

\beq H(\vec{k}) =  H_0 ( \vec{k} ) + H_w(\vec{k}) = v_{F} \tau^{z} ( \hat{z} \times \vec{\sigma}) .  (\vec{k} - k_{B_x} \tau^{z} \hat{y} -  k_{B_y} \tau^{z} \hat{x}) + \Delta_t \tau^x +  \frac{\lambda}{2} (k_+^3 + k_-^3) \sigma^z .\label{m}\eeq

Now, conveniently $[z,k_x] =[z,k_y]=[k_x,k_y]=0$, and so we can see that the combined Hamiltonian of eqns.(1) \& (2) has momentum eigenstates. Exploiting momentum commuation relations,
\beq k_+^3 + k_-^3 = 3k_x (k_x^2 - 3k_y^2),\eeq
and since the $k$ components have been gauge transformed as in eqn.(3), we arrive at
\beq k_+^3 + k_-^3 = K_1 + K_ 2 \tau^z  ,\eeq 
with $K_1$ and $K_2$ being defined as folllows:
\beq K_1 :=v_F^3[ k_x^3 + 3k_x k^2_{B_x} -3 k_x k_y^2 - 3k_x k^2_{B_y} + 6k_{B_x} k_{B_y} k_y] \eeq
\beq K_2 := v_F^3[ 3k_{B_x} k^2_y + 3k_{B_y}^2 k_{B_x} - k^3_{B_x} + 3k_{B_x} k_x^2 + 6k_{B_y} k_y k_x] .\eeq

The combined Hamiltonian H in (\ref{m}) can be written as

\beq H = 
\left [ \begin{smallmatrix}
\frac{\lambda}{2}(K_1 +K_2)   && v_F(k_y + ik_x -k_{B_x} +ik_{B_y}) && \Delta_t && 0\\
 v_F(k_y - ik_x -k_{B_x} -ik_{B_y}) &&-\frac{\lambda}{2}(K_1 +K_2 )  && 0 && \Delta_t \\
\Delta_t && 0 &&  \frac{\lambda}{2}(K_1 -K_2)   &&  v_F(-k_y - ik_x -k_{B_x} +ik_{B_y})  \\
0 && \Delta_t &&  v_F(-k_y + ik_x -k_{B_x} -ik_{B_y})  &&  -\frac{\lambda}{2}(K_1 -K_2 )  \\
\end{smallmatrix}  \right 
]
\eeq

Now, it so happens that we can diagonalize this Hamiltonian straightforwardly, and the four band dispersion are as follows:
\begin{multline}
 E^{s}_{\pm} = \pm \frac{1}{\sqrt{2}} [ 2v_F^2 k^2 + 2v_F^2 k_B^2 + 2\Delta_t^2 +  {\lambda^2}K_1 K_2   - s[\lambda^4 K_1^2 K_2^2 + 4\lambda^2 \Delta_t^2 K_1^2 +16 v_F^4 (k_y k_{B_x} - k_x k_{B_y})^2 + \\  8 v_F^2 \lambda^2 K_1 K_2 (k_x k_{B_y} - k_y k_{B_x}) +16 v_F^2 \Delta_t^2 (k_{B_x}^2 + k_{B_y}^2)]^{1/2}  ]^{1/2}
\end{multline}

where $s = \pm$ indicate the two sets of upper and lower bands. Shown below is the trivial case of $\lambda =0$, which is consistent with ~\cite{zyuin1103}, and it will prove useful in the next section.

\beq
 E^{s}_{\pm} = \pm  [ v_F^2 k^2 + v_F^2 k_B^2 + \Delta_t^2  - 2s[ v_F^2 \Delta_t^2 k_{B_x}^2 + v_F^4 k_y^2 k_{B_x}^2  -2 v_F^4 k_x k_y k_{B_x} k_{B_y} +  v_F^2 \Delta_t^2 k_{B_y}^2 +v_F^4 k_x^2 k_{B_y}^2    ]^{1/2}  ]^{1/2} 
\eeq

These bands which are symmetric in $k_x$ and $k_y$ and in all subsequent plots, $k_y$ has been set equal to zero, and the reason behind this will be stated shortly. Also, we assume $B=2T$ and $\phi =0$, except in Fig. 4.
\newline

\begin{figure}
\fbox{	\includegraphics[width=5.5 cm]{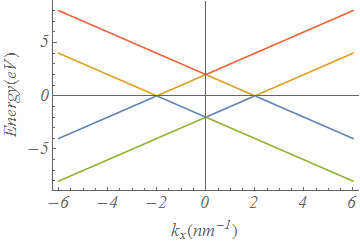}}
	\hspace*{1cm}
\fbox{	\includegraphics[width=5.5 cm]{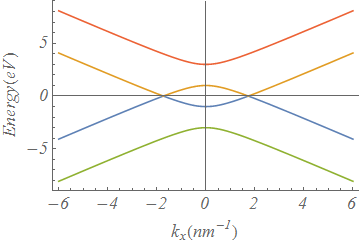}}
	%\hspace*{1cm}
	%\includegraphics[width=5.5 cm]{1bl}	
	\caption{Left: E vs $k_x$ with $\Delta_t=0$, and $\lambda=0$ Right: E vs $k_x$ with $\Delta_t = 1 meV$, and $\lambda=0$} 
\end{figure}

\begin{figure}
\fbox{	\includegraphics[width=5.5 cm]{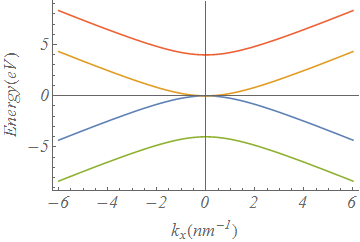}}
	\hspace*{1cm}
\fbox{	\includegraphics[width=5.5 cm]{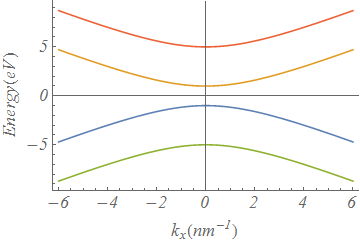}}
	%\hspace*{1cm}
	%\includegraphics[width=5.5 cm]{1bl}	
	\caption{Left: E vs $k_x$ with $\Delta_t>\epsilon_B$, and $\lambda=0$ Right: E vs $k_x$ with $\Delta_t=\epsilon_B = 1 meV$, and $\lambda=0$} 
\end{figure}

\begin{figure}
\fbox{	\includegraphics[width=5.5 cm]{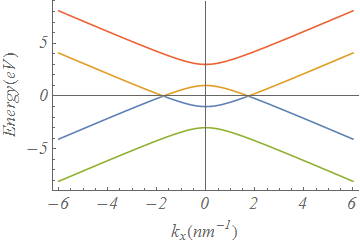}}
	\hspace*{1cm}
\fbox{	\includegraphics[width=5.5 cm]{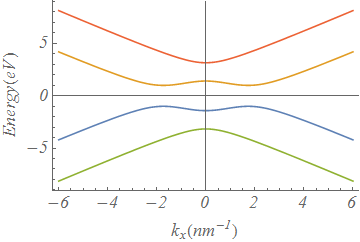}}
	%\hspace*{1cm}
	%\includegraphics[width=5.5 cm]{1bl}	
	\caption{Left: E vs $k_x$ with $\Delta_t = 1 meV$, and $\lambda=0.01 eV-nm^3$ Right: E vs $k_x$ with $\Delta_t = 1 meV$, and $\lambda=1 eV-nm^3$} 
\end{figure}

\begin{figure}
\fbox{	\includegraphics[width=5.5 cm]{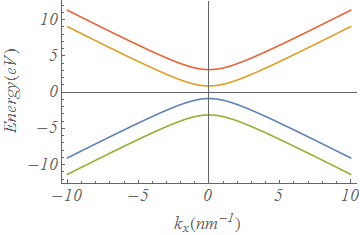}}
	\hspace*{1cm}
\fbox{	\includegraphics[width=5.5 cm]{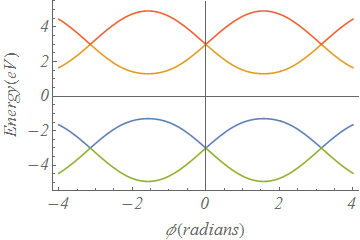}}
	%\hspace*{1cm}
	%\includegraphics[width=5.5 cm]{1bl}	
	\caption{Left: E vs $k_x$ with $\Delta_t=1 meV$, $\phi=0.6 rad$, and $\lambda=0$ Right: E vs $\phi$ with $\Delta_t = 1 meV$, $k_x = 2 nm^{-1}$ and $\lambda=0$} 
\end{figure}

The plots shown in Fig. 1 and Fig. 2 encapsulate the physics already discussed in ~\cite{zyuin1103}, with the Fermi energy $\epsilon_F =0$. In the limit of zero hybridization, the Fig. 1 (left) shows a semimetal state with four band crossing points, two of which are Dirac points (at ${\cal E}=0$). Usually surface electrons are unaffected by the presence of an inplane magnetic field which can be demonstrated by a suitable gauge transformation of the electron field operator. The introduction of a finite tunneling matrix element (with $\Delta_t < \epsilon_B = v_F k_B = v_F k_{B_x}$, since the field is only in the x direction) breaks this gauge symmetry and leads to the removal of two of these crossing points while moving the Dirac points around in momentum space as shown in Fig. 1 (right). Fig. 2 (left) shows the critical point of phase transition when $\Delta_t = \epsilon_B$, where the Dirac points merge leading to a vortex-antivortex like annihilation~\cite{Burkov}. When $\Delta_t > \epsilon_B$ the Dirac points have been removed and an insulating phase as shown in Fig. 2 (right) is obtained. 
\newline
\newline
We consider the effect of a small warping parameter on the semimetal state dispersion as shown in Fig. 3 (left). This newly introduced term leads to the annihilation of the the Dirac points, which is clearly discerned by cranking up the parameter ($\lambda=0.01 eV-nm^3$ to $\lambda= 1 eV-nm^3$), as plotted in Fig. 3 (right). It shows that when the Dirac points which are no longer protected against topological disorder in the presence of an inplane magnetic field~\cite{Burkov, Volo}, they are subject to efficient removal by introducing a warping contribution (which also leads to a spherical asymmetry of the Fermi surface) in conjunction with the hybridization piece. Thus a phase transition may be engineered by suitably altering the hybridization of an ultrathin TI with a given hexagonal warping parameter, which may be useful to the design of nano-devices. Now, a little consideration leads one to realize that the hexagonal warping term breaks chiral symmetry (similar to the tunneling term), and can be used in the absence of hybridization to achieve the aforementioned phase transition~\cite{Kane10}. This effect vanishes once the magnetic field has been turned off ~\cite{Kane10, Burkov}. It so happens that the Dirac points are stable to intersurface hybridization even for topologically unprotected edge state electrons as long as they are separated in momentum space, but this does not hold true for hexagonal warping which removes the unprotected Dirac points irrespective of location, as shown by Fig. 3.
\newline
\newline
Fig.4 (left) shows the band structure for a non-zero value of the tilt angle of the magnetic field. The introduction of a tilted magnetic field, similar to having a finite $k_y$  leads to positive definite contributions to the dispersion and a consequent increase in opening of the band gap. Finally, Fig. 4 (right) shows the variation of the band structure with the tilt angle, and the dispersion is periodic in $\pi$ which represents the physical description of invariance under discrete rotations by the stated angle. It is clear that an interplay of both tilt angle (except $\phi= 0, {\pi}, {2\pi}$) and warping leads to a global surface insulating phase.

\section{Hamiltonian due to a perpendicular electric field and Topological Phase Transition}
\subsection{Effects of an Electric Field in the limit of vanishing hybridization}

 We now propose the introduction of a perpendicular (to $x-y$ plane) electric field, so that our Hamiltonian in Sec. II is modified as follows. Consider a constant electric field $\vec{{\cal E}} = {\cal E}_0 \hat{z}$, such that the corresponding scalar potential is $\Phi = |e| {\cal E}z = \frac{|e| {\cal E}d}{2} \tau^z $. The modified Hamiltonian assumes the form:

\beq H = 
\left [ \begin{smallmatrix}
 \frac{\lambda}{2}(K_1 +K_2)  + \frac{|e| {\cal E}d}{2} &&  v_F(k_y + ik_x -k_{B_x} +ik_{B_y}) && \Delta_t && 0\\
 v_F(k_y - ik_x -k_{B_x} -ik_{B_y}) && -\frac{\lambda}{2}(K_1 +K_2 ) +\frac{|e| {\cal E}d}{2}  && 0 && \Delta_t\\
\Delta_t && 0 &&  \frac{\lambda}{2}(K_1 -K_2)   -\frac{|e| {\cal E}d}{2} &&  v_F(-k_y - ik_x -k_{B_x} +ik_{B_y})  \\
0 && \Delta_t &&  v_F(-k_y + ik_x -k_{B_x} -ik_{B_y})  && -\frac{\lambda}{2}(K_1 -K_2 ) - \frac{|e| {\cal E}d}{2} \\
\end{smallmatrix}  \right 
]
\eeq

Before examining the physics of the combined system with intersurface tunneling, it is more insightful to consider the influence of a pure electric field first so that the component wise effects are well documented. In doing so, if we naively set $\Delta_t =0$ with the assumption that the hybridization or "amount of communication between the two ends of the TI" can be reduced (by changing $d$), the eigenvalues are obtained as,

\beq 
E^{1}_{\pm} = \frac{|e| {\cal E}d}{2} \pm \sqrt{ \frac{\lambda}{2}^2( K_1 +K_2)^2 + v_F^2 [ k_x^2 + k_y^2 + k_{B_x}^2 + k_{B_y}^2 - 2k_x k_{B_y} +2k_y k_{B_x}]} 
\eeq
\beq
E^{2}_{\pm} = -\frac{|e| {\cal E}d}{2} \pm \sqrt{ \frac{\lambda}{2}^2( K_1 - K_2)^2 + v_F^2 [ k_x^2 + k_y^2 + k_{B_x}^2 + k_{B_y}^2 + 2k_x k_{B_y} -2k_y k_{B_x}]} ,
\eeq

where, the $1$ and $2$ superscripts indicate the two sets of upper and lower bands. 
\newline
\newline
Unfortunately, it turns out that the dispersion relations obtained fail to agree with Fig. 1 (left) in the vanishing electric field limit ($ {\cal E}=0$). This can be traced back to the fact that while it is entirely possible to first determine the eigen-energies with a defined tunneling element which can be set to zero later, it is not possible to reverse the ordering of these two processes. From a physical perspective, this could be interpreted as the fact that even though we assume a zero hybridization limit, there exists remenants of such an interaction within the dispersion which is irreversible. In fact, it turns out that the correct expression for zero electric field is obtained by setting $\Delta_t=0$ in eqn.(12).

For the case of a finite electric field, the exact solution can be obtained by determining the roots of the depressed quartic characteristic equation of the Hamiltonian. However, in this section, we pursue the numerical solutions and corresponding band structure plots of this Hamiltonian while persisting in the limit of zero tunneling between top and bottom surfaces. We examine the effects of introducing an electric field as shown in Fig. 5 (right) and note that the Dirac points start moving vertically in momentum space which leads to a part of the conduction band being lowered below the Fermi energy. This suggests a possible phase transition from a semimetal to a metallic state even though there is no visible overlap of conduction and valence bands. It also clear that since the introduction of the electric field coupled with the pseudospin does not break chiral symmetry or translational invariance (the latter can also lead to the hybridization of the Dirac cones), the Dirac points remain topologically protected~\cite{Burkov, Kane10} and cannot be annihilated by further increasing the field strength as  shown in Fig. 6 (left). It can be shown that on increasing the electric field from zero leads to the Dirac points approaching each other until a critical least separation point is reached beyond which they start to move apart again. Thus an insulating phase cannot be obtained from the semimetal TI surface by the use of an electric field in combination with an inplane magnetic field for an ultrathin TI.

The switching on of a small hexagonal warping facilitates the decoupling of the Dirac points and consequentially a phase transition to an insulator is obtained [Fig. 6 (right)] as already discussed.

\begin{figure}

\fbox{\includegraphics[width=5.5 cm]{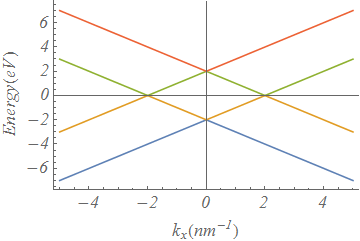}}
\hspace{1cm}
\fbox{\includegraphics[width=5.5 cm]{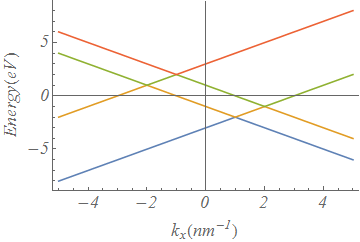}}
\caption{Left:E vs $k_x$ with $\Delta_t =0$, $ {\cal E}=0$, and $\lambda=0$, Right: E vs $k_x$ with $\lambda=0$ and $\Phi = 1 eV$} 

\end{figure}

\begin{figure}

\fbox{\includegraphics[width=5.5 cm]{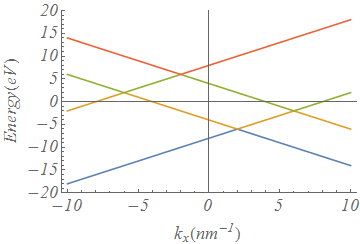}}
\hspace{1cm}
\fbox{\includegraphics[width=5.5 cm]{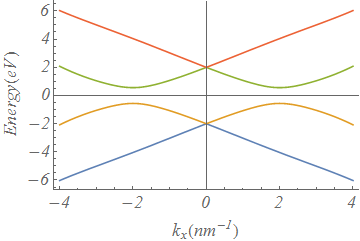}}
\caption{Left:E vs $k_x$ with $\lambda=0$ and $\Phi = 5 eV$  Right:E vs $k_x$ with $\lambda=0.1 eV-nm^3$ and $\Phi = 1 eV$} 

\end{figure}

\subsection{Combined Effects of Electric Field and Hybridization}

We now look at the effect of considering both an electric field and a non-zero tunneling element on an ultrathin TI subject to an inplane magnetic field in the $x$-direction. This is motivated by the fact that we now have exemplified the effects of an electric field and the tunneling element independently. As stated in the previous section we take a numerical approach to the problem at hand and then make some observations using established topological results~\cite{Volo}. Switching on a hybridization energy of $\Delta_t = 1 meV$ breaks "gauge" symmetry and again this leads to the movement of the Dirac points closer together in momentum space as shown in Fig. 7 (top left). In fact, as shown in Fig. 7 (top right), for a suitably chosen $\Delta_t = 1.732 meV $, we  happen upon a merger and subsequent anihilation of Dirac points (Fig. 7 (bottom)), i.e. a semi-metal to insulator phase transition. 
\newline

\begin{figure}

\fbox{\includegraphics[width=5.5 cm]{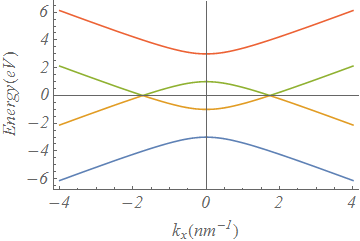}}
\hspace{1cm}
\fbox{\includegraphics[width=5.5 cm]{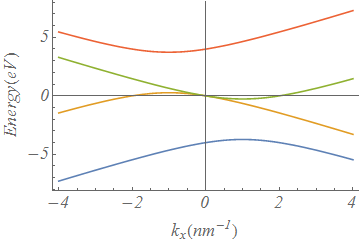}}
\vspace{1cm}
\hspace{1cm}
\fbox{\includegraphics[width=5.5 cm]{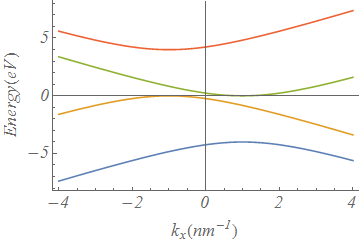}}

\caption{Top Left::E vs $k_x$ with $\lambda=0$, $\Delta_t =1 MeV$ and $\Phi = 4 eV$, Top Right: E vs $k_x$ with $\lambda=0$, $\Delta_t= 1.732 meV$ and $\Phi = 4 eV$ \newline Bottom: E vs $k_x$ with $\lambda=0$, $\Delta_t= 2 meV$ and $\Phi = 4 eV$ } 

\end{figure}

\begin{figure}

\fbox{\includegraphics[width=5.5 cm]{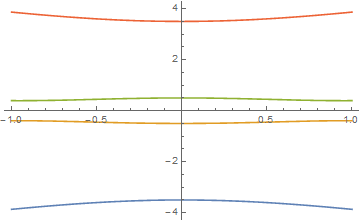}}
\caption{E vs $k_x$ with $\lambda=0.1 eV-nm^3$, $\Delta_t= 1 meV$ and $\Phi = 4 eV$} 

\end{figure}

It is noteworthy that the phase transition no longer occurs at $\Delta_t = \epsilon_B$, but is now non-trivially dependent on the electric field. The expression that relates the relevant quantities at the point of phase transition is found to be
 
\beq \epsilon_B^2 - \frac{{\cal E} ^2e^2d^2}{4} = \Delta_t^2 , \eeq

where ${\cal E}$ is the electric field, and this is plotted in Fig. 7 (bottom) with $d = 1 nm$. In the limit of zero electric field this reduces to the established condition for the critical point of QPT, i.e. $\epsilon_B = \Delta_t$. Eqn.(17) is in concurrence with ~\cite{Volo}, wherein Volovik and Klinkhamer have examined similar Dirac fermion systems within the context of CPT violations, exhibiting symmetry breaking and phase transitions parallel to the discussed in the present analysis. This leads to the notion that the set of all gapped bands form an equivalence class (all bands in this class are diffeomorphic to each other) and have an identical topological invariant (Chern number/ winding number) which means they are topologically equivalent to the vacuum. The surface states of the TI are gapless, which stems from having a different topological invariant characterized by the presence of the Dirac points. Thus phase transition from the semimetal phase to an insulating phase siginifies their equivalence class with the gapped bands and is thus a topological phase transition.
\newline
\newline
Finally, in Fig. 8 (right) we demonstrate the effects of including a non vanishing warping term, which predictably leads to an instantaneous separation of Dirac points and consequential semimetal-insulator phase transition as discussed before. We also note that on increasing $\Delta_t$ in this case simply leads to opening up the band gap further.

\section{Conclusion}

In this work, we have examined the effects of a confluence of different external fields and intersurface tunelling on a ultrathin warped 3D topological insulating film. The application of in plane tilted magnetic field causes topological disorder in the surface Dirac fermions and on turning on a hybridization element, a semimetal to insulator phase transition is obtained as a function of $\frac{\epsilon_B}{\Delta_t}$~\cite{Burkov}. We show that this effect can be reproduced independently by including a chiral symmetry breaking hexagonal warping term which destroys the Dirac points of the semimetal on occurence; this is different from the annihilation of the Dirac points by an vortex-antivortex like mechanism exhibited solely in the presence of the hybridization term. The band structure of the TI has also been scrutinized in the presence of a finite $|k_y|$ and an insulating phase has been found as a result, explained by the band gap increasing contribution of $k_y$ to the dispersion. The variation of the band structure with the tilt angle of the magnetic field has also been considered and the corresponding energy curves have been found to be periodic in phases of $\pi$ which is consistent with the geometry of the TI film. \\

On introducing a constant electric the Dirac points become vertically displaced as seen in the band structure; but this predictably does not induce a phase transition since the scalar potential does not break ${\cal PT}$ symmetry. Chiral symmetry can now be broken by introducing a finite tunneling matrix element which does not lead to immediate phase transition since Dirac points which are separated in momentum space are not susceptible to intersurface hybridization. The Dirac points approach each other on increasing the hybridization strength until a critical point is reached. As noted previously, the point of phase transition in the presence of the electric field is no longer at $\Delta_t = \epsilon_B$, but is now coupled to the electric field as stated in Eqn.(15): a concurrent result was obtained in a different context by ~\cite{Volo}. Subsequently, this phase transition has been briefly discussed using a topological perspective ~\cite{Kane10}, whereby the set of gapped bands form an equivalence class charcterized by the same Chern invariant. Finally, we introduced a small hexagonal warping term which causes spontaneous symmetry breaking driven semimetal-insulating phase transition; it is noted that this QPT can be achieved in the absence of the hybridization element. \\

{\bf  Acknowledgement}: Anirudha Menon would like to thank Prof. A. A. Zhyuzin (Department of Physics and Astronomy, University of Waterloo) for a correspondence that helped clarify a few points made in~\cite{zyuin1103}. 
The authors would also like to thank the anonymous referee whose insightful commentary on the first draft lead to the final version of this manuscript.

\end{document}